\documentclass[useAMS,usenatbib,psfig]{mn2e}
\usepackage{graphicx,rotating,times}
\usepackage{longtable,supertabular,multirow}
\usepackage{amsmath,amssymb,rotating}
\usepackage{lscape}
\newcommand{\hii}{{H{\scriptsize II} }}

\newcommand{\nhone}{\mbox{NH$_3$(1,1)}}

\newcommand{\thebrick}{\mbox{G0.253$+$0.016}}

\def \kms{\mbox{kms$^{-1}$}}

\def \cm2{\mbox{cm$^{-2}$}}
\def \cm3{\mbox{cm$^{-3}$}}

\title[SSC progenitor clouds in the GC]{Candidate super star cluster
  progenitor gas clouds possibly triggered by close passage to Sgr A*}

\author[S.N.Longmore et al.]{\parbox{\textwidth}{S.~N.~Longmore$^{1,2}$\thanks{E-mail: \texttt{S.N.Longmore@ljmu.ac.uk}}, 
J.~M.~D.~Kruijssen$^{3}$, 
J.~Bally$^{4}$, 
J.~Ott$^{5}$, 
L.~Testi$^{1,6}$, 
J.~Rathborne$^{7}$,
N.~Bastian$^{2}$,
E.~Bressert$^{7}$, 
S.~Molinari$^{8}$, 
C.~Battersby$^{4}$, 
A.~J.~Walsh$^{9}$} \vspace{0.4cm}  \\
$^{1}$European Southern Observatory, Karl-Schwarzschild-Strasse 2, D-85748 Garching bei M\"{u}nchen, Germany\\
$^{2}$Astrophysics Research Institute, Liverpool John Moores University, Twelve Quays House, Egerton Wharf, Birkenhead CH41 1LD\\
$^{3}$Max-Planck Institut fur Astrophysik, Karl-Schwarzschild-Strasse 1, 85748, Garching, Germany \\
$^{4}$Center for Astrophysics and Space Astronomy, University of Colorado, UCB 389, Boulder, CO 80309\\
$^{5}$National Radio Astronomy Observatory, P.O. Box O, 1003 Lopezville Road, Socorro, NM 87801, USA\\
$^{6}$INAF-Osservatorio Astrofisico di Arcetri, Largo E. Fermi 5, I-50125 Firenze, Italy\\
$^{7}$CSIRO Astronomy and Space Science, Epping, Sydney, Australia \\
$^{8}$INAF-Istituto Fisica Spazio Interplanetario, Via Fosso del Cavaliere 100, I-00133 Roma, Italy\\
$^{9}$International Centre for Radio Astronomy Research, Curtin University, GPO Box U1987, Perth WA 6845, Australia
\\
}

\begin{document}

\date{Accepted 2013 April 2. Received 2013 March 25; in original form 2013 February 19}
\pagerange{\pageref{firstpage}--\pageref{lastpage}} \pubyear{2013}
\label{firstpage}
\maketitle


\begin{abstract}

    Super star clusters are the end product of star formation under
    the most extreme conditions. As such, studying how their final
    stellar populations are assembled from their natal progenitor gas
    clouds can provide strong constraints on star formation
    theories. An obvious place to look for the initial conditions of
    such extreme stellar clusters are gas clouds of comparable mass
    and density, with no star formation activity. We present a method
    to identify such progenitor gas clouds and demonstrate the
    technique for the gas in the inner few hundred pc of our
    Galaxy. The method highlights three clouds in the region with
    similar global physical properties to the previously identified
    extreme cloud, $\thebrick$, as potential young massive cluster
    (YMC) precursors. The fact that four potential YMC progenitor
    clouds have been identified in the inner 100\,pc of the Galaxy,
    but no clouds with similar properties have been found in the whole
    first quadrant despite extensive observational efforts, has
    implications for cluster formation/destruction rates across the
    Galaxy. We put forward a scenario to explain how such dense gas
    clouds can arise in the Galactic centre environment, in which YMC
    formation is triggered by gas streams passing close to the minimum
    of the global Galactic gravitational potential at the location of
    the central supermassive black hole, Sgr~A*. If this triggering
    mechanism can be verified, we can use the known time interval
    since closest approach to Sgr~A* to study the physics of stellar
    mass assembly in an extreme environment as a function of
    \emph{absolute} time.

 \end{abstract}


\section{Introduction}

Stars in the most massive and dense stellar clusters must form at high
protostellar densities and in close proximity to large numbers of
high-mass stars. The dynamical interactions and (proto)stellar
feedback they experience make this one of the most extreme
environments in which stars can form. The progenitor clouds of these
super star clusters therefore provide an ideal laboratory for
understanding how environmental conditions affect the physics
governing star formation.

The inner few hundred parsecs of the Milky Way -- the `Central
Molecular Zone' (CMZ) -- is an ideal location in the Galaxy to search
for molecular cloud progenitors of the most massive
($>$10$^{4-5}$\,M$_\odot$) and dense (radius $\sim$1\,pc) stellar
clusters \citep[often called young massive clusters (YMCs);
see][]{portegieszwart2010}. The CMZ holds a substantial molecular gas
reservoir of $2 - 7 \times 10^7$\,M$_\odot$
\citep{morris_serabyn1996,ferriere2007} which has an average volume
density two orders of magnitude larger than that in the disk, and it
has been known for several decades that parts of this gas reservoir
contains very cold, dense cores with little signs of star formation
activity \citep[e.g the ``dust ridge'': ][]{lis1994, lis1998, lis1999,
  lis2001}.

Several studies have highlighted one CMZ molecular cloud in particular
-- variously know as M0.25, $\thebrick$, the Brick or the Lima Bean --
as extreme and potentially representing the initial conditions of a
YMC \citep{lis1998, lis2001, bally2010, longmore2012_brick}. In
retrospect, $\thebrick$ was easy to identify because it is so bright
and isolated in the far-IR/sub-mm emission maps, and stands out so
clearly as an IR absorption feature. However, $\thebrick$ contains
less than one hundredth of the total mass of the CMZ. It is therefore
possible that other, less conspicuous, YMC progenitor clouds may exist
but have not yet been identified as such in previous work. In this
Letter we return to the same data that \citet{longmore2012_brick} used
to identify and characterise $\thebrick$, but now attempt a more
systematic approach to finding YMC progenitor clouds in the CMZ.

\section{Identifying molecular cloud precursors of bound YMCs in the Galactic centre}

Most lines of sight along the Galactic plane contain substantial
emission from gas at varying distances. However, the vast majority of
the far-IR and sub-mm continuum emission within $|l|$$\sim$1$^\circ$
is from molecular gas at the Galactic centre (GC) distance
\citep{morris_serabyn1996, molinari2011}. Uniform angular resolution
then relates directly to a uniform physical resolution and uniform
flux sensitivity corresponds to a roughly uniform mass
sensitivity. Under these circumstances, it is then possible to
calculate the total mass as a function of radius from any given pixel
in a column density map, and use this to identify YMC progenitor
clouds.

However, from the column density map alone it is not possible to tell
how much of the mass within that radius is physically associated --
i.e. there is no guarantee that the projected distance in the plane of
the sky bears any relation to the physical separation between two
points. Also, given the complicated velocity structure in the CMZ, a
single pixel may contain flux contributions from multiple,
physically-distinct components along the line of sight. However, this
is easily identified by referring to molecular line data where the
additional velocity information uncovers the gas kinematic structure.

Even if the gas kinematics shows the emission at one spatial position
is from a single object, it is possible that the object may be
significantly more extended along the line of sight than inferred from
the projected radius in the plane of the sky. The average volume
density would therefore be lower than that assuming spherical
symmetry. It is possible, for instance, that high column density peaks
observed towards the GC may be elongated filaments (like those seen in
the barred spiral galaxy NGC 1097) seen end-on. We use the threshold
volume density, $n_{thresh}$, required for gas clouds to overcome the
extreme tidal forces at a distance $R_{GC}$ from the GC
($n_{thresh}>10^4$\,cm$^{-3} \times (75$\,pc$/R_{GC}) ^{1.8}$
\citep{guesten_downes1980}) to estimate the extent of clouds along the
line of sight. For clouds at representative distances from the GC of
50 and 100\,pc \citep{ferriere2007, molinari2011}, this implies a
maximum radius for spherical gas clouds massive enough to form YMCs
($\sim$10$^5$\,M$_\odot$) of approximately 1 to 3\,pc. By choosing
these projected radii limits to define the ``enclosed mass'' below,
any gas clouds must have a similar extent along the line of sight as
they do in the plane of the sky, otherwise they would quickly be
shredded into tenuous gas.

\citet[][hereafter B12]{bressert2012b} propose that bound YMCs form
from massive ($\gtrsim$10$^5$\,M$_\odot$) clouds enclosed within a
sufficiently small radius that the escape speed exceeds the sound
speed in photo-ionised gas. These criteria infer an additional
criteria of the minimum mass as a function of radius required for a
gas cloud to proceed to form a bound YMC.

Using the HiGAL column density map as a measure of the spatial
distribution of mass \citep{molinari2011}, and HOPS molecular line
data to resolve the kinematic structure \citep{walsh2011,purcell2012},
we now aim to assess which parts of the CMZ pass the B12 criteria, and
hence identify candidate YMC progenitor clouds.

\begin{figure*}
\begin{center}
\includegraphics[height=0.95\textwidth, angle=-90, trim=0 0 -5 0]{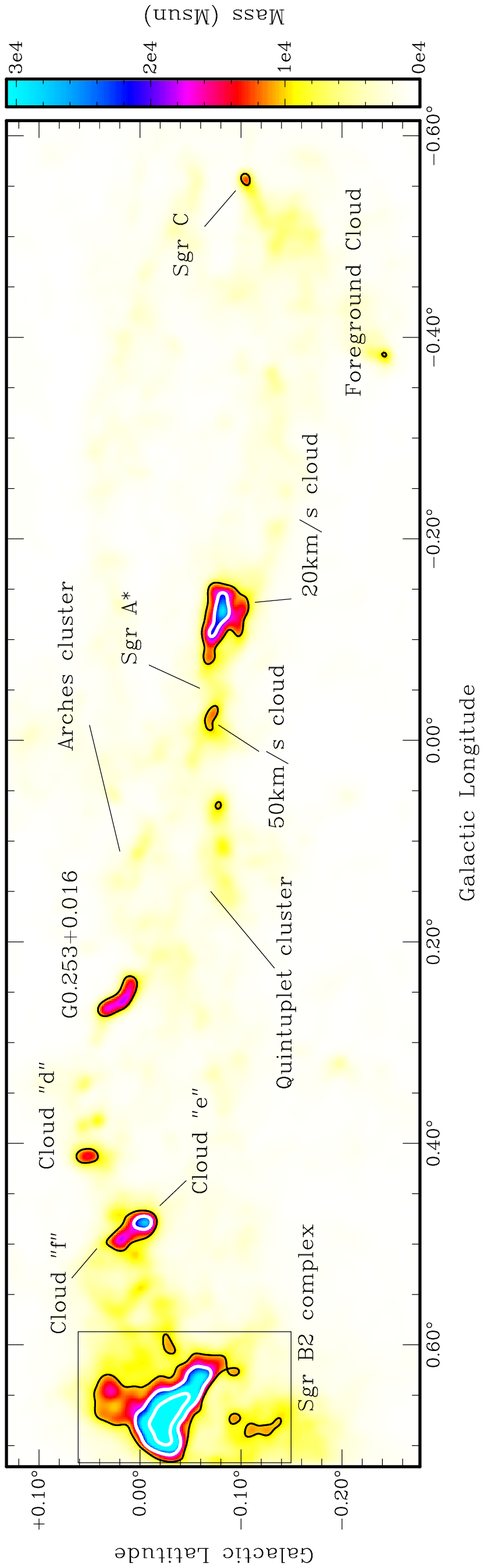} \\
\includegraphics[height=0.95\textwidth, angle=-90, trim=0 0 -5 0]{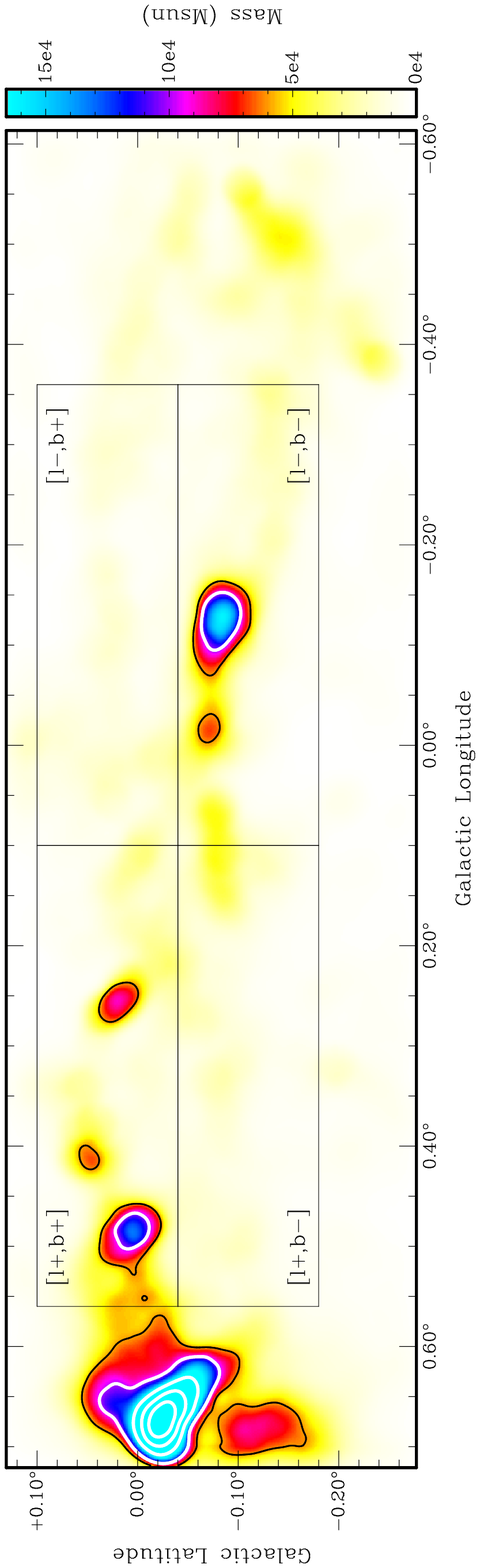} \\
\end{center}
\caption{\small{Maps of the ``enclosed-mass'' as a function of
    projected radius for the 100\,pc ring orbiting the centre of the
    Milky Way. Every pixel in the image shows the mass within a
    projected physical radius of 1\,pc [top] and 3\,pc [bottom] of
    that pixel, derived from the HiGAL column density map of the
    region \citep[][]{molinari2011, battersby2011}. The contours in
    the top image are at an enclosed mass within 1\,pc of 1, 2 and
    5$\times$10$^4$\,M$_\odot$. The contours in the bottom image are
    at enclosed mass within 3\,pc of 1, 2, 3, 4 and 5
    $\times10^5$\,M$_\odot$. The scale bar on the right panel shows
    the linear mass colour stretch. The lowest contours in each panel,
    shown in black rather than white, correspond roughly to the mass
    vs radius criteria in \citet{bressert2012b} required for a gas
    cloud to form a bound YMC.}}
\label{fig:enclosed_mass}
\end{figure*}

We do this on a pixel-by-pixel basis across the column density map by
calculating the mass enclosed within a given projected radius on the
plane of the sky. The top and bottom panels of
Figure~\ref{fig:enclosed_mass} show the resulting ``enclosed-mass''
map for a projected radius of 1\,pc and 3\,pc, respectively. The
colour scale reflects the mass range, the magnitude of which is given
by the colour bar on the right-hand edge of the panel. The lowest
(black) contour levels in each panel show the approximate threshold
mass within 1\,pc [top] and 3\,pc [bottom] that B12 predict should
form a bound YMC. The regions enclosed within these contours are
initial candidate YMC progenitor clouds.

The top panel of Figure~\ref{fig:enclosed_mass} contains annotations
identifying each of the candidate YMC progenitor clouds. The location
of the central, supermassive black hole, Sgr~A*, and the Arches and
Quintuplet clusters are also shown for orientation. This area of the
Galaxy has been studied intensively, so these sources are generally
well known \citep[see][]{morris_serabyn1996}. Sgr~B2 and Sgr~C host
\hii~regions, and Sgr~B2's high star formation rate means it is often
referred to as a mini-starburst. The 20 and 50\,$\kms$ clouds lie
closest in projection to the supermassive black hole, Sgr~A*, and it
has been suggested they are currently being tidally disrupted by a
close interaction \citep{herrnstein_ho2005}. $\thebrick$ and Clouds
``d'', ``e'' and ``f'' are well-studied, sub-mm-bright sources in the
so-called ``dust ridge'' \citep[see][for
  details]{lis1994,lis1998,immer2012}.

The 1\,pc enclosed-mass map (top panel of
Figure~\ref{fig:enclosed_mass}) shows one candidate source close to
the B12 limit which is not a well-known object, at ($l$,$b$) $\sim$
($-0.39$, $-0.25$). This is easy to identify in the HOPS $\nhone$ data
cubes as a foreground cloud, by the much narrower linewidth (a few
$\kms$) compared to the rest of the clouds in the CMZ. The mass
determined from the column density map assuming a GC distance is
therefore an overestimate. We remove this source from the list of
candidate YMC precursors.

The other candidate at ($l$,$b$) $\sim$ ($0.1$, $-0.05$) shows only a
single velocity component at this position in the HOPS $\nhone$ data
and the broad linewidth is consistent with this gas lying at the GC
distance. While this candidate is close to the B12 criteria at an
enclosed-mass radius of 1\,pc (top panel of
Figure~\ref{fig:enclosed_mass}), it is not above the B12 criteria at
an enclosed-mass radius of 3\,pc (bottom panel of
Figure~\ref{fig:enclosed_mass}). We therefore do not include this as a
robust candidate YMC precursor cloud in further analysis. A similar
argument holds for Sgr~C. The 20 and 50\,$\kms$ clouds are likely to
be distinct physical objects, but their close passage to Sgr~A* makes
their dynamical state uncertain, so we do not include them as YMC
precursor candidates. Sgr~B2 is known to have a complex kinematic
structure with multiple velocity components as a function of position
and several distinct and physically separated star formation regions
\citep[see][and references therein]{qin2011}. The assumptions used to
calculate the enclosed-mass in Figure~\ref{fig:enclosed_mass}
therefore do not hold for this region.

This leaves $\thebrick$ -- a previously-identified YMC progenitor --
and clouds ``d'', ``e'' and ``f''.  \citet{immer2012} derive the
detailed properties of clouds ``d'', ``e'' and ``f''. They have masses
of 7.2, 15.3 and 7.2$\times$10$^4$\,M$_\odot$, and radii of 3.5, 4.5
and 2.7\,pc, respectively. HOPS $\nhone$ data shows they have very
similar integrated line profiles to $\thebrick$, and virial analysis
shows them to be close to gravitational stability. As a final check,
we independently searched the ATCA NH$_3$ GC survey data (Ott et al in
prep.) and confirmed these sources as bright, isolated, compact, dense
gas peaks with a single velocity component at higher angular
resolution.

In summary, we reconfirm $\thebrick$ as a potential YMC progenitor
cloud using this method and highlight three other clouds (``d'', ``e''
and ``f'') which also pass the B12 criteria. 

\section{Implications for YMC formation}

Further observations to derive the detailed gas properties of each
cloud are required to determine if these candidate progenitors will
form a YMC. In particular, observations\footnote{For example, direct
  volume density measurements \citep[e.g.][]{ginsburg2011} or
  comparison to numerical models \citep[e.g.][]{clark2013}} are needed
to determine if the assumption of approximately spherical clouds is
valid. This may not be the case. Dust structures and lanes in
circum-nuclear gas rings of galaxies often show filamentary
``streamers'' \citep[see e.g.][]{peeples_martini2006}. If such
streamers exist in our own Galaxy, and they have orientations similar
to those in external galaxies, one would expect to find them
``end-on'' to our line of sight in the receding velocity part of the
first quadrant. This is where we find all the highest column density
gas and candidate YMC progenitor clouds. If some of these are
filamentary streamers, they may be on the verge of being sheared by
tides -- a natural explanation of their lack of star formation
\citep[see e.g.][]{kauffmann2013}.

However, we know that YMCs form in this part of the
Galaxy, so we should also expect to find precursor clouds. The clouds
highlighted above are the best candidates for the initial conditions
of future Arches-like stellar clusters. Therefore, it seems reasonble
to assume that at least one of these will proceed to form such a
cluster.

If YMCs all form in a similar way from clouds with similar initial
conditions, comparing the number of clouds at the same evolutionary
stages in different locations provides a direct comparison of the YMC
formation rate between the regions. Therefore, the fact that
\citet{ginsburg2012} find no starless YMC progenitor clouds with
similar mass and density in the first quadrant of the Galaxy ($6^\circ
< l < 90^\circ$, $|b|<0.5^\circ$) implies that more YMCs are currently
forming per unit time in the GC than the disk. For the ISM conditions
in the inner few hundred pc of the GC, \citet{kruijssen2012} predicts
a much higher fraction ($\sim$50\%) of stars will form in
gravitationally-bound stellar clusters compared to the local solar
neighbourhood ($\sim$7\%). So per unit star formation rate the CMZ
should be much more efficient at forming bound clusters.  However,
YMCs show no preferred galacto-centric radius in any galaxy, including
the Milky Way \citep{portegieszwart2010}. Reconciling these facts
would require that the YMCs forming in the centre of the Milky Way
have correspondingly shorter lifetimes after formation than elsewhere
in the Galaxy. Simulations suggests this is the case
\citep[e.g.][]{portegieszwart2001, kruijssen2011}. In the central 300
pc of their model galaxy, \citet{kruijssen2011} find the disruption
time is at least an order of magnitude shorter than in the disk. These
model predictions are consistent with the large number of observed
progenitor molecular clouds in the CMZ compared to the rest of the
Galaxy but no correspondingly large number of long-lived YMCs. In this
scenario, while more are being created, they are also being destroyed
at a higher rate, so at any given snapshot in time the YMC number
density does not vary with galacto-centric radius.

Another possibility is that the formation mechanism for YMCs outside
the GC is different. Rather than forming in a short burst from the
prompt collapse of such dense gas clouds, YMCs may form over an
extended period much longer than the free-fall time, via continuous
accretion \citep[e.g.][]{smith2009}. In this case one would not expect
to see $\thebrick$-like clouds in the Galactic disk. The implications
of this -- that YMC formation mechanisms vary with environment -- has
important consequences for interpreting observed YMC distributions in
external galaxies.

\subsection{YMC formation triggered by gas interacting with the
  gravitational potential around Sgr~A*?}
\label{sub:trigger?}

In an attempt to distinguish between these possibilities, we now try
to understand how the CMZ environment may be playing a role in
creating such massive and dense molecular clouds that appear to be
about to form YMCs. 

We note that the distribution of the \emph{densest} gas in
Figure~\ref{fig:enclosed_mass} is asymmetric. The potential YMC
progenitor clouds lie at positive latitudes between roughly Sgr~A* and
Sgr~B2. To see if this asymmetry in gas density is reflected in the
distribution of total mass, we broke up the region in to four
rectangular segments of equal area, split at half the projected
distance between Sgr~B2 and Sgr~C. We discarded Sgr~B2 and Sgr~C
themselves due to saturation and potential line of sight issues. The
$l$ and $b$ ranges of the four areas are listed in
Table~\ref{tab:regions_mass}, and shown on the bottom panel of
Figure~\ref{fig:enclosed_mass}. The final column in
Table~\ref{tab:regions_mass} shows the total mass in each of the four
quadrants, which all agree to within a factor of two. The increased
density in the $[l+,b+]$ quadrant is not simply due to a higher total
mass.

\begin{figure}
\begin{center}
\includegraphics[width=0.45\textwidth, angle=0, trim=0 0 -5 0]{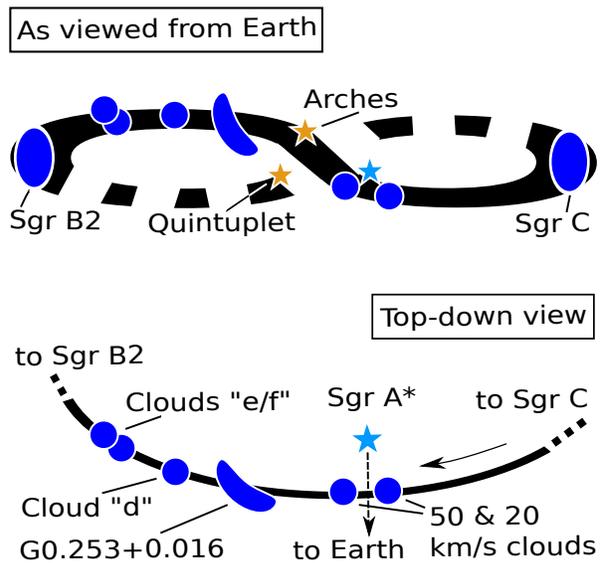} \\
\end{center}
\caption{\small{[Upper] Schematic diagram of the gas in the inner
    $\sim$1$^\circ$ of the Galaxy as viewed from Earth.  The thick
    solid and dashed black lines represent the stream of gas moving
    from Sgr~C to Sgr~B2 and the far side of the ring, respectively,
    in the \citet{molinari2011} model. [Lower] Top-down view of this
    gas stream. The curved arrow shows the sense of rotation. Between
    Sgr~C and Sgr~B2 the gas passes close to the bottom of the
    Galactic gravitational potential denoted by the supermassive black
    hole, Sgr~A*. The upper and lower parts are approximately aligned
    vertically. }}
\label{fig:schematic}
\end{figure}

We then seek an explanation for what might be causing the increased
gas density. In order to do this, we need to make some assumptions
about the 3D geometry of the gas. \citet{molinari2011} recently
proposed that the molecular gas within $\sim$1$^\circ$ of the GC lies
in a ring orbiting the GC. An interesting aspect of this model is that
the supermassive black hole, Sgr~A*, is not at the geometric
centre. Instead, it is closer to the front side of the ring. As the
schematic diagram in Fig~\ref{fig:schematic} shows, the gas therefore
passes close to the bottom of the Galactic gravitational potential as
it orbits from Sgr~C to Sgr~B2. It seems reasonable to assume the gas
may have been affected by the varying gravitational potential along
this orbit, with the strongest affect at pericentre passage. Note that
the location of Sgr~A* relative to the gas is important in as far as
it represents the bottom of the Galactic gravitational
potential. However, Sgr~A*'s radius of gravitational influence is
$\lesssim$2\,pc.  Therefore, given plausible gas trajectories, the
gravitational field felt by the gas is likely to be dominated instead
either by the nuclear cluster surrounding Sgr~A* or the nuclear
stellar disk, which are the main contributors to the potential at
radii of 2~--~30\,pc and 30~--~300\,pc, respectively
\citep{launhardt2002}.

Gas moving on an orbit around a source of strong gravitational
potential will feel a combination of two effects as it approaches
pericentre. Firstly it will experience increasing compression in the
vertical direction perpendicular to the orbit. At the same time it
will also become more stretched along the orbit. It is interesting to
note in this regard that $\thebrick$, a cloud in this scenario that
recently passed through pericentre to the bottom of the Galactic
potential, appears extended along the orbit proposed by
\citet{molinari2011} and has a small scale height compared to the
majority of the gas in the ring.

The fate of the gas after pericentre passage will depend on many
factors. We are currently investigating this scenario numerically,
paying particular attention to understanding the phase space
distribution of the gas (Kruijssen, Dale, Longmore et al. in prep) and
comparing this directly to observations (Rathborne, Longmore et al. in
prep).

In this scenario, one interpretation of the density contrast of gas up
and downstream from pericentre passage, is that the net effect of the
interaction is a compression of the gas. We speculate that this is
aided by the gas dissipating the tidally injected energy, which would
be observable as strongly shocked gas. A direct prediction of this is
that the hydrodynamic shocks should be strongest around pericentre
passage than elsewhere in the region.

How might the interaction with the bottom of the Galactic
gravitational potential affect the star formation activity? Given the
reservoir of dense gas available to form stars, the gas in the inner
few hundred pc of the Galaxy is known to be under-producing stars by
at least an order of magnitude compared to commonly-assumed star
formation relations \citep{longmore2013_sfrgc}. So if the gas was
previously sitting close to gravitational stability, the additional
net compression of the gas might be enough for it to begin collapsing
to form stars. With this scenario in mind, it is interesting to note
that the YMC progenitor clouds progressively farther down stream from
pericentre passage with Sgr~A* show progressively more star formation
activity. The closest cloud downstream, $\thebrick$, shows little
signs of star formation activity \citep[e.g.][]{longmore2012_brick,
  kauffmann2013, rodriguez_zapatat2013}. Cloud ``d'' has methanol
maser emission, signposting massive star formation is underway
\citep{immer2012}. As mentioned above, Sgr~B2 has prodigious star
formation activity.

Further observations are required to test this tentative evolution of
star formation activity from Sgr~A* to Sgr~B2. However, if the
hypothesis proves correct, the implications are potentially
exciting. Given the observed oribital velocity of the gas, we can
calculate the time since each of the clouds passed pericentre i.e. the
time at which star formation may have been instigated. Assuming an
orbital velocity of 80\,$\kms$ \citep{molinari2011}, the projected
distances of $\thebrick$, Cloud ``d'' and Clouds ``e/f'' from Sgr~A*
suggest they passed pericentre approximately 0.6, 0.9 and 1.0\,Myr
ago, respectively.  Therefore, we may have a truly unique opportunity
to effectively follow the physics shaping the formation of the most
massive stellar clusters in the Galaxy, and by inference the next
generation of the most massive stars in the Galaxy, as a function of
\emph{absolute} time.

An additional prediction of this scenario is that the resulting
stellar clusters should have kinematics consistent with their having
formed from gas on this orbital trajectory. As shown in
Figure~\ref{fig:schematic}, the two known YMCs in the region -- the
Arches and Quintuplet clusters -- are observed to lie in projection
along the ring. However, the cluster ages are comparable to the ring's
orbital period, so their current locations do not necessarily reflect
where they formed. In addition, determining the true distance of these
clusters from the GC and constraining their orbital properties is
observationally challenging \citep[][]{figer1999, figer2002,
  hussmann2012, stolte2008, clarkson2012}. While some observed
properties (e.g. the cluster proper motions) are consistent with them
having been associated with the gas in the ring, further work is
needed to determine if the proposed scenario can explain the origin of
the Arches and Quintuplet clusters.

We emphasise that the scenario outlined above depends on the model of
\citet{molinari2011} in only two ways. Firstly, we assume that the gas
moving from Sgr~C to Sgr~B2 is a coherent stream. Secondly, we assume
that between these two points the gas passes close to the bottom of
the Galactic gravitational potential. The scenario does not rely on
other aspects of the \citet{molinari2011} model, such as the rotation
velocity, whether Sgr~B2 is closer/farther from Earth than Sgr~A* or
whether Sgr~B2/Sgr~C are tangent points of the ring at the
intersection of the $x_1$ and $x_2$ orbits.

\subsection{Asymmetry in the total gas mass distribution?}

Returning to the values in Table~\ref{tab:regions_mass}, it is
striking that the mass for the $[l+,b+]$ and $[l-,b-]$ quadrants agree
so closely, as do the $[l+,b-]$ and $[l-,b+]$ quadrants. However, the
mass of the former are a factor of two larger than the latter. Could
this be the result of a systematic bias? The gas in this region is all
thought to lie within $\sim$100\,pc of the GC
\citep[e.g.][]{ferriere2007, molinari2011} so the distance-dependence
on the mass is at the ($\Delta$distance/distance)$^2\sim(100/8500)^2
\sim 10^{-4}$ level so can be neglected. While there may be mechanisms
through which the dust properties may vary in the GC environment, it
would be curious that such a disparity has not been noticed before in
such a well-observed region. We note that the $[l+,b+]$ and $[l-,b-]$
quadrants and $[l+,b-]$ and $[l-,b+]$ quadrants encapsulate the near
and far sides of the ring, respectively, in the \citet{molinari2011}
model. It is interesting to speculate whether the apparent factor of
two decrease in gas mass on the far side of the ring may be either
caused by changes in dust properties, or gas leaving the ring due to
star formation activity within Sgr~B2. We flag this as an interesting
avenue for further investigation.

\section{Conclusions}

We present a method for finding YMC progenitor clouds and use this to
identify four candidate proto-YMC clouds in the inner few hundred pc
of our Galaxy. We discuss the significance of finding four such YMC
progenitor clouds with very little signs of star formation in such a
small volume of the Galaxy, while no starless YMC progenitor clouds
have been found in the first quadrant of the Galaxy. We infer that in
environments like the Galactic centre, YMCs either form via a
different mechanism, or the formation and destruction times are much
shorter. We then investigate the distribution of the gas and put
forward a scenario to explain the observed asymmetry in the dense gas
distribution. We propose that gas is compressed by passing close to
the minimum of the global Galactic gravitational potential. We
speculate that this may instigate the condensation of dense YMC
progenitor clouds, which leads to the subsequent formation of stars
towards Sgr~B2. If this hypothesis can be verified we may have a truly
unique opportunity to effectively follow the physics shaping the
formation of the most massive stellar clusters in the Galaxy, and by
inference the next generation of the most massive stars in the Galaxy,
as a function of \emph{absolute} time.

We would like to thank the anonymous referee for their prompt,
positive and constructive comments. SNL would like to thank Henrik
Beuther, Ian Bonnell, Jim Dale, Betsy Mills, Mark Morris and Malcolm
Walsmley for insightful discussions. SNL acknowledges this research
has received funding from the European Community's Seventh Framework
Programme (/FP7/2007-2013/) under grant agreement No 229517.

\begin{table}
\begin{center}
  \caption{Properties of the four equal-area regions discussed in the
    text and illustrated in the bottom panel of
    Figure~\ref{fig:enclosed_mass}.}
\label{tab:regions_mass}
\begin{tabular}{ccccccc}\hline

Region       & $l_{min}$ & $l_{max}$ & $b_{min}$ & $b_{max}$ & Mass \\
             & [deg.]   & [deg.]  & [deg.]   & [deg.]   & [10$^5$\,M$_\odot$] \\ \hline   
1. $[l+,b+]$ & 0.1      & 0.56    & -0.04    &  0.1     & 11.9\\
2. $[l+,b-]$ & 0.1      & 0.56    &-0.18     & -0.04    & 6.1\\
3. $[l-,b+]$ & -0.36    & 0.1     & -0.04    & 0.1      & 5.6 \\
4. $[l-,b-]$ & -0.36    & 0.1     & -0.18    & -0.04    & 11.2\\ \hline


\end{tabular}
\end{center}
\end{table}

\bibliography{isolated_gc_clumps}

\begin{thebibliography}{}
\small
\bibitem[\protect\citeauthoryear{{Bally}, {Aguirre}, {Battersby}, {Bradley} \&
  {Cyganowski}}{{Bally} et~al.}{2010}]{bally2010}
{Bally} J.,  {Aguirre} J.,  {Battersby} C.,  {Bradley} E.~T.,    {Cyganowski}
  C.,  2010, \apj, 721, 137

\bibitem[\protect\citeauthoryear{{Battersby}, {Bally}, {Ginsburg} \&
  {Bernard}}{{Battersby} et~al.}{2011}]{battersby2011}
{Battersby} C.,  {Bally} J.,  {Ginsburg} A.,    {Bernard} J.-P.,  2011, \aap,
  535, A128

\bibitem[\protect\citeauthoryear{{Bressert}, {Ginsburg}, {Bally} \&
  {Battersby}}{{Bressert} et~al.}{2012}]{bressert2012b}
{Bressert} E.,  {Ginsburg} A.,  {Bally} J.,    {Battersby} C.,  2012, \apjl,
  758, L28

\bibitem[\protect\citeauthoryear{{Clark}, {Glover}, {Ragan}, {Shetty} \&
  {Klessen}}{{Clark} et~al.}{2013}]{clark2013}
{Clark} P.~C.,  {Glover} S.~C.~O.,  {Ragan} S.~E.,  {Shetty} R.,    {Klessen}
  R.~S.,  2013, ArXiv e-prints

\bibitem[\protect\citeauthoryear{{Clarkson}, {Ghez}, {Morris}, {Lu}, {Stolte},
  {McCrady}, {Do} \& {Yelda}}{{Clarkson} et~al.}{2012}]{clarkson2012}
{Clarkson} W.~I.,  {Ghez} A.~M.,  {Morris} M.~R.,  {Lu} J.~R.,  {Stolte} A.,
  {McCrady} N.,  {Do} T.,    {Yelda} S.,  2012, \apj, 751, 132

\bibitem[\protect\citeauthoryear{{Ferri{\`e}re}, {Gillard} \&
  {Jean}}{{Ferri{\`e}re} et~al.}{2007}]{ferriere2007}
{Ferri{\`e}re} K.,  {Gillard} W.,    {Jean} P.,  2007, \aap, 467, 611

\bibitem[\protect\citeauthoryear{{Figer}, {McLean} \& {Morris}}{{Figer}
  et~al.}{1999}]{figer1999}
{Figer} D.~F.,  {McLean} I.~S.,    {Morris} M.,  1999, \apj, 514, 202

\bibitem[\protect\citeauthoryear{{Figer}, {Najarro}, {Gilmore}, {Morris},
  {Kim}, {Serabyn}, {McLean}, {Gilbert}, {Graham}, {Larkin}, {Levenson} \&
  {Teplitz}}{{Figer} et~al.}{2002}]{figer2002}
{Figer} D.~F.,  {Najarro} F.,  {Gilmore} D.,  {Morris} M.,  {Kim} S.~S.,
  {Serabyn} E.,  {McLean} I.~S.,  {Gilbert} A.~M.,  {Graham} J.~R.,  {Larkin}
  J.~E.,  {Levenson} N.~A.,    {Teplitz} H.~I.,  2002, \apj, 581, 258

\bibitem[\protect\citeauthoryear{{Ginsburg}, {Bressert} \& {Bally}}{{Ginsburg}
  et~al.}{2012}]{ginsburg2012}
{Ginsburg} A.,  {Bressert} E.,    {Bally} J.,  2012, \apjl, 758, L29

\bibitem[\protect\citeauthoryear{{Ginsburg}, {Darling}, {Battersby}, {Zeiger}
  \& {Bally}}{{Ginsburg} et~al.}{2011}]{ginsburg2011}
{Ginsburg} A.,  {Darling} J.,  {Battersby} C.,  {Zeiger} B.,    {Bally} J.,
  2011, \apj, 736, 149

\bibitem[\protect\citeauthoryear{{Guesten} \& {Downes}}{{Guesten} \&
  {Downes}}{1980}]{guesten_downes1980}
{Guesten} R.,  {Downes} D.,  1980, \aap, 87, 6

\bibitem[\protect\citeauthoryear{{Herrnstein} \& {Ho}}{{Herrnstein} \&
  {Ho}}{2005}]{herrnstein_ho2005}
{Herrnstein} R.~M.,  {Ho} P.~T.~P.,  2005, \apj, 620, 287

\bibitem[\protect\citeauthoryear{{Hu{\ss}mann}, {Stolte}, {Brandner}, {Gennaro}
  \& {Liermann}}{{Hu{\ss}mann} et~al.}{2012}]{hussmann2012}
{Hu{\ss}mann} B.,  {Stolte} A.,  {Brandner} W.,  {Gennaro} M.,    {Liermann}
  A.,  2012, \aap, 540, A57

\bibitem[\protect\citeauthoryear{{Immer}, {Menten} \& {Schuller}}{{Immer}
  et~al.}{2012}]{immer2012}
{Immer} K.,  {Menten} K.~M.,    {Schuller} F.,  2012, \aap, 548, A120

\bibitem[\protect\citeauthoryear{{Kauffmann}, {Pillai} \& {Zhang}}{{Kauffmann}
  et~al.}{2013}]{kauffmann2013}
{Kauffmann} J.,  {Pillai} T.,    {Zhang} Q.,  2013, ArXiv e-prints

\bibitem[\protect\citeauthoryear{{Kruijssen}}{{Kruijssen}}{2012}]{kruijssen201%
2}
{Kruijssen} J.~M.~D.,  2012, \mnras, 426, 3008

\bibitem[\protect\citeauthoryear{{Kruijssen}, {Pelupessy} \&
  {Lamers}}{{Kruijssen} et~al.}{2011}]{kruijssen2011}
{Kruijssen} J.~M.~D.,  {Pelupessy} F.~I.,    {Lamers} H.~J.~G.~L.~M.,  2011,
  \mnras, 414, 1339

\bibitem[\protect\citeauthoryear{{Launhardt}, {Zylka} \& {Mezger}}{{Launhardt}
  et~al.}{2002}]{launhardt2002}
{Launhardt} R.,  {Zylka} R.,    {Mezger} P.~G.,  2002, \aap, 384, 112

\bibitem[\protect\citeauthoryear{{Lis}, {Li}, {Dowell} \& {Menten}}{{Lis}
  et~al.}{1999}]{lis1999}
{Lis} D.~C.,  {Li} Y.,  {Dowell} C.~D.,    {Menten} K.~M.,  1999, in {Cox} P.,
  {Kessler} M.,  eds, The Universe as Seen by ISO Vol.~427 of ESA Special
  Publication, {Cold GMC cores in the Galactic Centre}.
p.~627

\bibitem[\protect\citeauthoryear{{Lis} \& {Menten}}{{Lis} \&
  {Menten}}{1998}]{lis1998}
{Lis} D.~C.,  {Menten} K.~M.,  1998, \apj, 507, 794

\bibitem[\protect\citeauthoryear{{Lis}, {Menten}, {Serabyn} \& {Zylka}}{{Lis}
  et~al.}{1994}]{lis1994}
{Lis} D.~C.,  {Menten} K.~M.,  {Serabyn} E.,    {Zylka} R.,  1994, \apjl, 423,
  L39

\bibitem[\protect\citeauthoryear{{Lis}, {Serabyn}, {Zylka} \& {Li}}{{Lis}
  et~al.}{2001}]{lis2001}
{Lis} D.~C.,  {Serabyn} E.,  {Zylka} R.,    {Li} Y.,  2001, \apj, 550, 761

\bibitem[\protect\citeauthoryear{{Longmore}, {Bally}, {Testi}, {Purcell} \&
  {Walsh}}{{Longmore} et~al.}{2013}]{longmore2013_sfrgc}
{Longmore} S.~N.,  {Bally} J.,  {Testi} L.,  {Purcell} C.~R.,    {Walsh} A.~J.,
   et al., 2013, \mnras, 429, 987

\bibitem[\protect\citeauthoryear{{Longmore}, {Rathborne}, {Bastian} \& {et
  al,}}{{Longmore} et~al.}{2012}]{longmore2012_brick}
{Longmore} S.~N.,  {Rathborne} J.,  {Bastian} N.,    {et al,} 2012, \apj, 746,
  117

\bibitem[\protect\citeauthoryear{{Molinari}, {Bally} \&
  {Noriega-Crespo}}{{Molinari} et~al.}{2011}]{molinari2011}
{Molinari} S.,  {Bally} J.,    {Noriega-Crespo} A.,  2011, \apjl, 735, L33

\bibitem[\protect\citeauthoryear{{Morris} \& {Serabyn}}{{Morris} \&
  {Serabyn}}{1996}]{morris_serabyn1996}
{Morris} M.,  {Serabyn} E.,  1996, \araa, 34, 645

\bibitem[\protect\citeauthoryear{{Peeples} \& {Martini}}{{Peeples} \&
  {Martini}}{2006}]{peeples_martini2006}
{Peeples} M.~S.,  {Martini} P.,  2006, \apj, 652, 1097

\bibitem[\protect\citeauthoryear{{Portegies Zwart}, {Makino}, {McMillan} \&
  {Hut}}{{Portegies Zwart} et~al.}{2001}]{portegieszwart2001}
{Portegies Zwart} S.~F.,  {Makino} J.,  {McMillan} S.~L.~W.,    {Hut} P.,
  2001, \apjl, 546, L101

\bibitem[\protect\citeauthoryear{{Portegies Zwart}, {McMillan} \&
  {Gieles}}{{Portegies Zwart} et~al.}{2010}]{portegieszwart2010}
{Portegies Zwart} S.~F.,  {McMillan} S.~L.~W.,    {Gieles} M.,  2010, \araa,
  48, 431

\bibitem[\protect\citeauthoryear{{Purcell}, {Longmore}, {Walsh}, {Whiting} \&
  {et al.}}{{Purcell} et~al.}{2012}]{purcell2012}
{Purcell} C.~R.,  {Longmore} S.~N.,  {Walsh} A.~J.,  {Whiting} M.~T.,    {et
  al.} 2012, \mnras, 426, 1972

\bibitem[\protect\citeauthoryear{{Qin}, {Schilke}, {Rolffs} \& {Comito}}{{Qin}
  et~al.}{2011}]{qin2011}
{Qin} S.-L.,  {Schilke} P.,  {Rolffs} R.,    {Comito} C.,  2011, \aap, 530, L9

\bibitem[\protect\citeauthoryear{{Rodriguez} \& {Zapata}}{{Rodriguez} \&
  {Zapata}}{2013}]{rodriguez_zapatat2013}
{Rodriguez} L.~F.,  {Zapata} L.,  2013, ArXiv e-prints

\bibitem[\protect\citeauthoryear{{Smith}, {Longmore} \& {Bonnell}}{{Smith}
  et~al.}{2009}]{smith2009}
{Smith} R.~J.,  {Longmore} S.,    {Bonnell} I.,  2009, \mnras, 400, 1775

\bibitem[\protect\citeauthoryear{{Stolte}, {Ghez}, {Morris}, {Lu}, {Brandner}
  \& {Matthews}}{{Stolte} et~al.}{2008}]{stolte2008}
{Stolte} A.,  {Ghez} A.~M.,  {Morris} M.,  {Lu} J.~R.,  {Brandner} W.,
  {Matthews} K.,  2008, \apj, 675, 1278

\bibitem[\protect\citeauthoryear{{Walsh}, {Breen}, {Britton} \& {et
  al.}}{{Walsh} et~al.}{2011}]{walsh2011}
{Walsh} A.~J.,  {Breen} S.~L.,  {Britton} T.,    {et al.} 2011, \mnras, 416,
  1764

\end{thebibliography}

\label{lastpage}

\end{document}